\documentclass[%
 reprint,
 amsmath,amssymb,
 aps,
]{revtex4-2}

\usepackage{graphicx}% Include figure files
\usepackage{dcolumn}% Align table columns on decimal point
\usepackage{mathrsfs}
\usepackage{xcolor}
\usepackage{bm}% bold math
\usepackage{atbegshi}
\AtBeginShipout{%
  \pdfpageattr{/Group << /S /Transparency /I true /CS /DeviceRGB >>}%
}

\begin{document}

\preprint{APS/123-QED}

\title{%A Novel Order-Reduction Technique to Study the Nonlinear Dynamics of Biomimetic Scale Metamaterials
Chaotic Flexural Vibrations in Biomimetic Scale Substrates}% Force line breaks with \\

\author{Omid Bateniparvar}
\author{Farzan Farahmand}
\author{Ranajay Ghosh}
\email{Ranajay.ghosh@ucf.edu}

 \affiliation{Department of Mechanical and Aerospace Engineering, University of Central Florida, Orlando, FL, USA}

\begin{abstract}
\noindent Overlapping fish-scale architectures are among nature's most distinctive surface adaptations, combining protection, contact regulation, hydrodynamics, optical and directional mechanical response within a thin textured integument. Here, we show that their biomimetic structural analogues can host deterministic chaos. Biomimetic scale substrates develop chaotic flexural vibrations at modest amplitudes because bending activates unilateral contact and progressive jamming, while built-in asymmetry from unequal texturing biases the restoring response and shifts the onset of chaos. From continuum mechanics, we derive a singular reduced-order model (sROM) that reduces the scale-covered beam to a nonlinear oscillator whose parameters map directly to overlap, scale inclination, damping, forcing, and substrate stiffness. Finite element (FE) simulations validate the model in quasi-static bending and long-time forced response. Stroboscopic regime maps reveal a period-doubling cascade from period-1 to period-2 and period-4, ultimately chaos. Overlap and inclination determine the strength of post-engagement nonlinearity, whereas damping bounds the chaotic operating window. Unequal top-bottom scale distributions break the antisymmetry of the restoring response, generating offset force-displacement laws. This reduced symmetry does not accelerate instability; instead, it delays the onset of chaos and fragments the response into intermittent periodic windows, whereas restoring symmetry can paradoxically widen the chaotic regime. When the texture is sufficiently sparse or steep on one side, it remains dynamically inactive, and the beam behaves as a fully asymmetric one-sided system. The results identify biomimetic scale substrates as a distinct class of contact-rich architectured metasurfaces in which chaos is programmable through geometry rather than large deflection or constitutive nonlinearity.\bigskip

\textbf{\textit{Significance:}}  Overlapping biological scales are an ancient integumentary adaptation, yet their influence on long-time forced dynamics of the underlying substrate has remained largely unexplored. We show that biomimetic scale substrates can generate deterministic chaos at modest amplitudes through unilateral contact and progressive jamming alone, without requiring large deflections or intrinsic material nonlinearity. A singular reduced-order model (sROM), validated against contact-resolved finite element (FE) simulations, reveals how overlap, scale inclination, damping, and top-bottom asymmetry select period-1, multi-period, or chaotic motion. In particular, damping contracts chaotic windows, whereas restoring symmetry can accelerate the onset of chaos. Conversely, broken symmetry can delay the onset of chaos and reduce the extent of chaotic response. These results elevate contact from a local irregularity to a programmable design mechanism for nonlinear vibration, with implications for impact management and adaptive mechanical response.\bigskip

\begin{center}
-----------------------------
\end{center}
\end{abstract}

\maketitle
\section{Introduction}
\label{Sec I}

%%%%%%%%%%%%%%%%%%%%%%%%%%%%%%%%%%%%%%%%%%%%%%

Overlapping scales are among nature’s most consequential surface architectures \textcolor{black}{\cite{sfakiotakis2002review,triantafyllou2005review,ghosh2014contact,hossain2022fish}}. In fishes and reptiles they combine protection, control of body–environment contact, and mechanically directional surface behavior in the same thin integument \textcolor{black}{\cite{chen2012predation,zhu2013puncture,chang2019evaluation}}. This makes them a rare biological example in which local geometry simultaneously governs load transfer, locomotor interaction, and fluid-mediated performance \textcolor{black}{\cite{lauder2016structure,chang2019evaluation,browning2013mechanics}}. Recent bioinspired work has therefore treated scale-covered substrates not simply as decorated beams, but as a mechanically active class of metasurfaces in which architecture and deformation are inseparable \textcolor{black}{\cite{ghosh2014contact,ali2019bending,ebrahimi2021fish,tatari2023bending}}. Yet the emphasis so far has been overwhelmingly quasistatic, such as quantifying strain-stiffening, fracture and penetration protection, drag-related surface effects, and material-geometry viscoelasticity \textcolor{black}{\cite{ali2019bending,tatari2020bending,ebrahimi2023material}}. More broadly, although periodic metabeams have been studied extensively in linear and weakly nonlinear regimes in the context of architected materials and metamaterials  \textcolor{black}{\cite{liu2000locally,zhang2015flexural}}, the dynamical consequences of contact in scale textured substrates remain unresolved. Here, most analyses emphasize band gaps and scattering \textcolor{black}{\cite{xiao2013flexural,nouh2015wave,zhang2019band,liu2020bandgaps}} while neglecting kinematic constraints, broken symmetry, friction, and contact sliding that potentially reorganize the underlying phase space \textcolor{black}{\cite{ghosh2016frictional,ali2019frictional,ebrahimi2020coulomb}}. Thus, long-time forced dynamics of such contact-textured beams remain largely unresolved.

What makes these systems distinctive is that their nonlinearity does not arise primarily from large geometric deflection or from constitutive complexity, but from the abrupt onset of feature engagement. Once neighboring scales begin to interact, bending is converted into a piecewise, geometry-dependent restoring response with broken antisymmetry, progressive hardening, and an eventual locking barrier \textcolor{black}{\cite{ghosh2014contact,ali2019bending,tatari2020bending}}. That mechanism is qualitatively different from the smooth nonlinearities that dominate most beam models, suggesting that contact-textured beams may host routes to complexity unavailable to conventional architected structures \textcolor{black}{\cite{sun2010theory,findeisen2017characteristics,kim2022nonlinear}}. Although the fish-scale architecture is the canonical instance of this overlapping-surface motif, the same underlying jamming mechanism is expected in substrates bearing discrete protrusions or filaments, such as spines, hairs, or fur, wherever unilateral contact generates comparable nonlinear and asymmetric restoring moments \textcolor{black}{\cite{krsmanovic2023fur}}.

Here, we show that contact nonlinearity alone is sufficient to produce organized transitions from near-harmonic motion to asymmetric limit cycles and deterministic chaos in scale-textured beams. Starting from continuum mechanics, we derive a reduced description of bending rigidity that can capture the three primary regimes of deformation: linear, nonlinear, and jammed \textcolor{black}{\cite{ghosh2014contact}}. The regimes are determined by overlap kinematics, periodic self-contacts, Coulomb friction, and substrate compliance \textcolor{black}{\cite{ghosh2016frictional,ali2019frictional,ebrahimi2020coulomb}}. This leads to a compact nonlinear oscillator model whose effective parameters map directly to measurable geometric and material quantities, including curvature, overlap ratio, scale inclination, substrate modulus, and damping \textcolor{black}{\cite{ebrahimi2021fish,ebrahimi2023material,sarkar2025bending}}. Fully resolved finite element (FE) simulations with explicit contact validate that the reduced model reproduces both time histories and phase portraits.
The proposed contact-driven model exposes clear routes to complex oscillations and chaotic vibrations without invoking large deflections. We show that contact discontinuity alone is enough to generate period-n motions and chaotic vibrations, with greater contact stiffness leading to a higher number of oscillation periods. Assuming scales with different distributions on the two sides, the model shows a co-evolution of both temporal oscillation complexity and accentuating phase-space asymmetry. We also find that, despite damping, contact nonlinearity, forcing frequency, and asymmetric geometry can interplay to preserve chaotic motions \textcolor{black}{\cite{kim2022nonlinear,bateniparvar2024dynamic,hossain2024biomimetic,fang2024advances}}. 

We map these chaotic regimes using the largest Lyapunov exponent (LLE) and Poincaré sections. A set of nondimensional groups collapses the parameter space, yielding design rules that link architectural choices to phase-space structure. Specifically, since contact stiffness is strongly determined by the sliding kinematics of the scales, embedding parameters of scales, and asymmetry of distribution, this chaos is primarily geometry-enabled.

This geometry-driven behavior, linked to the underlying elasticity of the substrate, enables the same framework to generalize to other textured metasurfaces that impose self-contact constraints, such as pillar arrays or interlocking shingles \textcolor{black}{\cite{khandelwal2015adaptive,shan2015multistable,karuriya2024fully,bateniparvar2026bioinspired}}. This establishes contact-induced nonlinear elasticity as a minimal mechanical primitive for programming vibration regimes in architected materials. The results enable geometry-tunable dissipation, oscillations, and impact energy management, and provide predictable operating windows where complex dynamics can be exploited rather than suppressed. 
Recent interest in soft robotic control has increasingly turned to physical reservoir computing \textcolor{black}{\cite{nakajima2020physical,bhovad2021physical,he2025physical}}, where useful computation arises from the intrinsic nonlinear dynamics of the body itself \textcolor{black}{\cite{mandal2022machine,he2025role}}. In that context, the vibrational regimes identified here may be relevant beyond mechanics alone. Specifically, by expanding the accessible repertoire of transient and nonlinear responses, contact-textured metamaterials could provide a geometrically tailorable platform for physical reservoirs \textcolor{black}{\cite{nakajima2020physical,zhang2022harnessing}}.

%%%%%%%%%%%%%%%%%%%%%%%%%%%%%
\begin{figure*}
\includegraphics[width=1\linewidth]{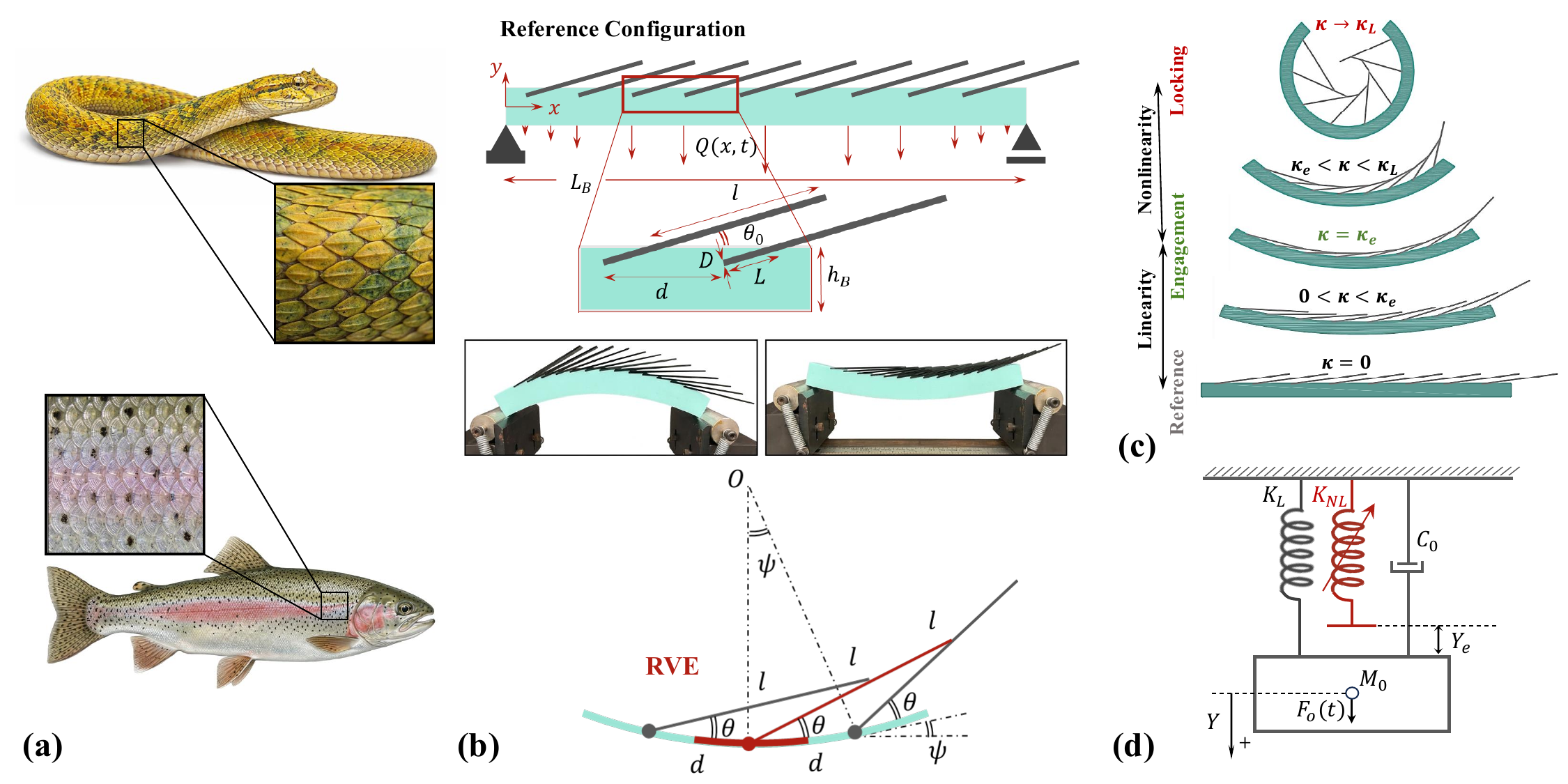}
\caption{\label{Fig1}\textbf{Overview of biological scales, biomimetic scale substrate and nonlinear spring-mass-damper (nSMD) system.} (a) Representative biological textures, including snake (\textit{Bothriechis schlegelii}, eyelash viper) and fish (\textit{Oncorhynchus mykiss}, rainbow trout), highlighting natural scale-based morphologies and surface functionality. Images are AI-generated for conceptual illustration. (b) Schematic of a biomimetic scale substrate with simply supported ends subjected to a spatially sinusoidal and temporally harmonic distributed load, $Q(x,t)$ (top), illustration of the system deformation in two opposite directions (middle), and the representative volume element (RVE) (bottom). (c) Pure bending of a biomimetic scale substrate, showing deformation progression from zero curvature ($\kappa=0$) through scale engagement ($\kappa=\kappa_e$) on to the locking curvature ($\kappa = \kappa_L$). (d) Equivalent nSMD system: during positive deflection, only the linear spring is active, whereas for negative deflection, after engagement ($Y\geq Y_e$), the nonlinear spring is engaged, capturing the asymmetric and nonlinear stiffness induced by scale contact.}
\end{figure*}

%%%%%%%%%%%%%%% METHOD SECTION 
\section{Methods and Models}
\label{Sec II}

We consider a slender beam of bending rigidity $B=E_BI_B$ where $E_B$ is the Young's modulus of the beam and $I_B$ is the cross-sectional moment of area. This beam is textured with rigid scales of length $l$ and scale spacing (pitch) $d$, initial inclination $\theta_0$, and overlap ratio $\eta=l/d$. Beam curvature $\kappa(x,t)$ ultimately leads to scale engagement and drives it into sliding. For $\kappa<\kappa_e(\eta,\theta_0)$ where $\kappa_e$ is the scale engagement curvature, there is no scale interaction, and the moment-curvature response is linear $M=B\kappa$. Beyond this point, the scales enter the engagement regime where the scales self-contact. Contact forces oppose the sliding motion. Each embedded scale is modeled as attached to the substrate through a torsional spring of stiffness $K_B$, which captures the base compliance \textcolor{black}{\cite{ghosh2014contact}}. The onset of engagement leads to a piecewise restoring force measured by the bending moment, which depends on the geometry and contact kinematics obtained by assuming periodic scales contact \textcolor{black}{\cite{ali2019bending,tatari2020bending,hossain2022fish}}. As the sliding moves deeper into engagement, eventually a jammed state is reached, characterized by a divergent tangent stiffness in the moment \textcolor{black}{\cite{ghosh2016frictional,ebrahimi2020coulomb,sarkar2025bending}} at the so-called locking curvature $\kappa_L$. At this stage,  deformation shifts from the bending of the softer substrate to the stiffer scale material. Dissipation sources can include external fluidic damping, interfacial Coulomb friction, and material viscosity. Even with Coulomb friction, biomimetic scale systems are known to exhibit emergent viscosity that tends to change most damping laws into viscous-like behavior \textcolor{black}{\cite{ali2019frictional,ebrahimi2020coulomb,ebrahimi2023material}}. Such viscous behavior can be modeled in rate form as proportional to curvature rate $\dot\kappa$.  Asymmetric textures brought on by different shapes and distribution of scales, \textcolor{black}{FIG.~\ref{Fig1}(b)}, can lead to asymmetric $M(\kappa)$. This asymmetry reflects differing nonlinear elasticity of the two sides due to the nature of the scales.

The mechanisms described above would require at each step, contact detection, contact constraint enforcement, resolution of various interdependencies between parameters for tangent stiffness calculations, and a progressively stronger nonlinearity. Thus, fully resolved scale substrate dynamical FE models tend to diverge without exceptionally small time steps. This can be prohibitively expensive, making parameter sweeps needed to traverse the entire landscape of nonlinear dynamics forbiddingly difficult. Thus, developing a singular reduced-order model (sROM) that can model the jamming behavior is of paramount importance. 

We first define a nondimensional unit cell rotation defined as $\psi=\kappa d$ (see \textcolor{black}{FIG.~\ref{Fig1}(b)}). We seek to develop a constitutive map $M(\kappa)=\partial_\kappa W(\psi;\mathbf{p})$ from curvature to bending moment via a reduced potential $W$ with the parameters $\mathbf{p}$ tied to overlap ratio, pitch, scales inclination, and material properties of the substrate. 
 
 This map must obey three major constraints that are defined by the characteristics of the bending process: (i) Linear onset $M=B\kappa$ for $|\kappa|<\kappa_e$ (pre-engagement), (ii) Engagement regime, $\kappa>\kappa_e$ with increasing tangent stiffness and (iii) Locking limit $\kappa \to \kappa_L$ where the tangent stiffness becomes singular in the rigid-scale limit, \textcolor{black}{FIG.~\ref{Fig1}(c)}. This three-phase barrier-type restoring law can be approximated at a unit-cell level as follows. First let $\psi_e$ be the engagement curvature and $\psi_L>\psi_e$ the locking curvature obtained from contact-kinematics under curvature control \textcolor{black}{\cite{ghosh2014contact,ali2019bending}}. We now define a scaled argument of type:

 \begin{align}
 \label{eq:1}
 &z(\kappa) = \frac{\pi}{2(\psi_L-\psi_e)}(\psi-\psi_e)
 \end{align}

This scaling maps the engagement interval $\psi \in [\psi_e, \psi_L]$ onto $z \in [0, \pi/2]$. Among candidate barrier functions, the tangent form provides a compact analytic representation that preserves linear onset, produces monotonic stiffening during engagement, and enforces divergence of the tangent stiffness at the locking limit.

\begin{align}
\label{eq:2}
&M_s(\kappa)=B\kappa+M_c(\kappa)\notag\\[8pt]
&M_c(\kappa)=H(z) A\eta^m \tan(\alpha z)
\end{align}

Here $\alpha$ is introduced as a regularization parameter to prevent singularities in numerical code and is generally close to but not unity. The constant $A$ represents nonlinear stiffness, and the power law $m$ on $\eta$ is appended to capture the strong overlap ratio dependence of moment-curvature behavior \textcolor{black}{\cite{ghosh2014contact,ali2019bending}}. The Heaviside function $H(\cdot)$ activates the nonlinear term only after scale engagement. In this work, it is regularized using a $\tanh$-based smooth approximation to ensure continuity~\textcolor{black}{\cite{he2015new}}. Specifically, we define $\tilde{H}(z) = \frac{1}{2}\left(1 + \tanh(\alpha z)\right)$, where $\tilde{H}(\cdot)$ is used to denote this continuous approximation.

This choice recovers the linear substrate for $\psi<\psi_e$ and produces a divergence at $\psi\rightarrow\psi_L$, matching the jamming limit. Note that $\tan z$ acts somewhat like a shape function, but we need the power law on overlap ratio to reproduce the entirety of the nonlinearity leading up to jamming.  

The associated nonlinear conservative potential for this system is denoted by $\Phi_{NL}$ and is a bounded function. Curvature parameters $\psi_e,\psi_L$ can be obtained from geometry of the contact kinematics \textcolor{black}{\cite{ghosh2014contact,ali2019bending}}. Asymmetry can be incorporated in this model by allowing the sROM parameters to differ for $\psi>0$ and $\psi<0$. $\psi_e$ is directly dictated by the overlap ratio $\eta$ and scales inclination $\theta_0$. This relationship, arising from the imposed periodicity of contact, can be expressed as $\eta\psi_e\cos(\psi_e/2)-\sin(\theta_0+\psi_e/2)=0$ \textcolor{black}{\cite{ghosh2014contact,ghosh2016frictional}}. This nonlinear bridging relationship between local scale rotation ($\theta$) and global unit cell rotation ($\psi$) achieves singularity at jamming (locking) kinematic limit $\psi_L\approx 1/\eta$  \textcolor{black}{\cite{ghosh2014contact}}.

The nonlinear stiffening parameters $A$, $m$, and regularization parameter $\alpha$ can be fitted with closed-form analytical moment-curvature relationships from existing literature \textcolor{black}{\cite{ghosh2014contact,ali2019bending,tatari2020bending,hossain2022fish}} where the scale rotation is modeled as equivalent torsional springs. The torsional spring constant \( K_B=0.66E_BD^2(L/D)^{1.75} \), and obtained by empirically fitting nondimensional groups with FE simulations \textcolor{black}{\cite{ghosh2014contact}}. Thus, as per the sROM construction, the coefficient \( A \) of the sROM is directly related to the torsional spring constant.

We now recast the beam dynamics in the small-slope Euler-Bernoulli kinematic limit $y' \ll 1$, so that  $ \kappa \approx y''$. Let $x\in (0,L_B)$ where $L_B$ is the beam length, and $Q(x,t)$ be the external transverse load \textcolor{black}{FIG.~\ref{Fig1}(b)}. The conservative Lagrangian density of the beam, including the nonlinear scale-contact potential $\Phi_{NL}$ can be written in the following form:

\begin{equation}
\label{eq:3}
    \mathcal{L}(y,\dot y,y'')=\frac{1}{2}\rho S\dot y^2-\left[ \frac{1}{2}B\kappa^2+\Phi_{NL}\right]+Qy, \quad \kappa\approx y''
\end{equation}

where the first term on the right-hand side is the kinetic energy density of the substrate quantified by the mass density $\rho$, the area of the cross section $S$, and the velocity field $\dot y$. In this calculation, we neglect both the rotary inertia (slender beam assumption) and the kinetic energy of the scales (lightweight-scale limit). However, they can be calculated from the kinematics of the underlying substrate bending. We observe that, under geometric constraints typical in various types of natural scales (e.g., $L/D>3, L/h_B<0.25, L/l<0.1$ \textcolor{black}{\cite{browning2013mechanics}}), even moderate $\eta$ yields $m_r\gg 1$ where $m_r$ refers to the substrate-to-scale mass ratio \textcolor{black}{\cite{ali2019frictional}}. Consequently, the mass of the substrate dominates the dynamics. The nonlinear contact potential $\Phi_{NL}$ reproduces the moment, $M_c=\partial \Phi_{NL}/\partial \kappa$ given in \textcolor{black}{Eq.~\ref{eq:2}} 
and rewritten for small deformation as: 
%%%%%%%%%%%%%%%%%%%%%%%%%%%%%

\begin{equation}
\label{eq:4}
M_{s}(y'')=B y''+\tilde{H}(y''-y''_e) A\eta^m  \tan\left( \frac{\pi}{2} \alpha \frac{y''-y''_e}{y''_L-y''_e} \right)
\end{equation}

where $y''_e$ and $y''_L$ represent the engagement and locking curvature, respectively. These curvatures are themselves geometrically predetermined. Dissipation of the fish-scale system is a complex process with contributions from both the underlying contact kinematics and frictional effects. However, as discussed above, prior studies confirming the emergent viscosity of this class of biomimetic beams allows us to define a dissipation function in a unified fashion to a leading order as $\mathcal R=\mathcal R(\kappa, \dot\kappa,\dot y)=\mathcal R(\dot y)=\frac{1}{2}c_d\dot y^2$ where $c_d$ is the damping coefficient.

Requiring stationarity of the action according to the extended Hamilton's principle with dissipation yields:

\begin{equation}
\label{eq:5}
\tilde{\delta} 
\int_{0}^{t}
\int_{0}^{L_B}
\mathcal{L}\bigl(y,\dot y,y''\bigr)
\,\mathrm{d}x\,\mathrm{d}t
+ 
\int_{0}^{t}
\int_{0}^{L_B}
\frac{\partial\mathcal{R}\bigl(\dot y\bigr)
}{\partial\dot{y}}\tilde{\delta}y\,\mathrm{d}x\,\mathrm{d}t=0
\end{equation}

where $\tilde{\delta}$ denotes the first (infinitesimal) variation operator. Considering both time and spatial derivatives yields the Euler–Lagrange equation as follows:

\begin{equation}
\label{eq:6}
\frac{\partial}{\partial t} \left( \frac{\partial \mathcal{L}}{\partial \dot{y}} \right) + \frac{\partial \mathcal{R}}{\partial \dot{y}}-\frac{\partial^2}{\partial x^2} \left( \frac{\partial \mathcal{L}}{\partial y''} \right)-\frac{\partial \mathcal{L}}{\partial y} = 0
\end{equation}

This leads to the following nonlinear partial differential equation (PDE):

\begin{equation}
\label{eq:7}
\rho S\ddot{y}+c_d\dot{y}+By^{iv}+\frac{\partial^2}{\partial x^2}
\left(\frac{\partial \Phi_{NL}(y'')}{\partial y''}\right)
= Q(x,t)
\end{equation}

We next rewrite the PDE above in terms of the restoring moment:

%%%%%%%%%%%%%%%%%%%%%%%%%%%%%%%%%%%%%%%%%
\begin{equation}
\label{eq:8}
\rho S\ddot{y}+c_d\dot{y}
+\frac{\partial^2 M_{s}(y'')}{\partial x^2}
= Q(x,t)
\end{equation}

According to \textcolor{black}{Eq.~\ref{eq:2}}, the governing equation is expressed in terms of the divergence
of the total bending moment, while the piecewise nature of contact is fully encoded within $M_{s}$.

%%%%%%%%%%%%%%%%%%%%%%%%%%%%%%%%%%%%%%%%%

For the forcing used in parametric sweeps we take time-harmonic load targeting the first mode under simply supported boundary conditions leading to $Q(x,t)=f_0\phi_1(x)\cos(\omega t)$ where $f_0$ is the amplitude of the applied load,  $\phi_1(x)$ is the first eigenmode of a simply supported beam (i.e.,  $\phi_1(x)=\sin(\pi x/L_B)$, see \textcolor{black}{FIG.~\ref{Fig1}(b)}).

The underlying substrate provides several intrinsic invariant normalization
quantities for the length and time scales.
These include the beam length for horizontal coordinate scaling,
i.e.\ $\bar{x} = x/L_B$; the beam height for vertical coordinate scaling,
i.e.\ $\bar{y} = y/h_B$, and the natural frequency for time and frequency scaling,
i.e.\ $\tau = \omega_0 t$ and $\Omega = \omega/\omega_0$. Here $\omega_0 = \pi^2 \sqrt{B/\rho S L_B^4}$ denotes the first natural frequency of the substrate for a simply supported beam \textcolor{black}{\cite{blevins2016natural,ali2019frictional}}.

After nondimensionalization, the nonlinear PDE (\textcolor{black}{Eq.~\ref{eq:8}}) can be converted to a nonlinear temporal ODE via a Galerkin weak-form approach and separation of variable $\bar{y}(\bar{x},\tau) = \phi_1(\bar{x}) Y(\tau)$ to yield the following nonlinear ODE for the time varying part (see \textcolor{black}{\ref{SM1}} for derivation details).

\begin{align}
\label{eq:9}
&\ddot{Y}(\tau)
+ \delta\,\dot{Y}(\tau)+\bar{F_s}=\gamma\,\cos\bigl(\Omega\,\tau\bigr) \notag\\[8pt]
&\bar{F_s}=Y(\tau)+\bar{F}_{NL}\bigl(Y(\tau), \bar{K}, \eta, \theta_0, \zeta, \bar{d}\bigr)\notag\\[8pt]
&\bar{K}=AL_B/B,\quad \zeta=h_B/L_B,\quad \bar{d}=d/L_B
\end{align}

In the above nondimensional ODE $\delta = c_d / \rho S \omega_0$, and
$\gamma = f_0 / \rho S h_B \omega_0^2$. $\bar{F}_{NL}$ represents the normalized nonlinear contact force generated by scale engagement and is a function of time and the geometric parameters governing the scale architecture (see \textcolor{black}{\ref{SM1}} for derivation details). Physically, increasing $\eta$ or decreasing $\theta_0$ promotes stronger contact interactions between neighboring scales. At larger $\eta$ or smaller $\theta_0$, scale engagement occurs earlier and more abruptly, leading to a significant amplification of the nonlinear contact force $\bar{F}_{NL}$. Note that \textcolor{black}{Eq.~\ref{eq:9}} represents a nondimensional nonlinear spring-mass-damper (nSMD) system (shown in \textcolor{black}{FIG.~\ref{Fig1}(d)}).\\

FE simulations were conducted using commercial FE software Abaqus (Dassault Systèmes) to analyze the dynamic response of a fully asymmetric biomimetic scale substrate, facilitating the validation of chaotic behavior. The substrate was modeled as a homogeneous elastic material, with material properties, geometric dimensions, and loading conditions detailed in \textcolor{black}{TABLE \ref{Tab:1}}. To ensure the generality of the observed dynamics, two distinct configurations (Case I and Case II) were considered, spanning different scale geometries and excitation regimes. While Case I represents a longer and more compliant configuration that serves as the primary reference for the sROM results discussed later, Case II corresponds to a shorter substrate with a higher excitation frequency. The scales were assumed to be rigid, with their density set to 5\% of the substrate density to minimize mass effects. Although no explicit physical damping was introduced in the FE model, the simulations are not strictly undamped. Commercial FE software, such as those used here contributes a small amount of numerical dissipation through bulk viscosity and time integration, so the FE results should be viewed as weakly dissipative rather than perfectly conservative. This is appropriate here because the FE model is used primarily to validate the contact-induced mechanics, while the role of damping is investigated separately and systematically in the sROM.

Here, default ABAQUS/Explicit values for linear and quadratic bulk viscosity parameters have been used. The beam was configured with simply supported boundary conditions. For Case I, the simulation employed a mesh of 10,491 elements (10,212 4-node plane strain elements with reduced integration (CPE4R), and 279 3-node plane strain elements (CPE3), whereas Case II employed 17,035 elements, including 16,781 CPE4R and 254 CPE3. These choices ensure a balance between computational efficiency and solution accuracy across both configurations \textcolor{black}{\cite{chen2018control,rocha2021numerical}}. 
A distributed oscillatory load, defined as \( Q = f_0 \sin(\pi x / L_B) \cos(\omega t) \), was applied to the biomimetic scale substrate, with parameters specified in \textcolor{black}{TABLE \ref{Tab:1}}. Contact interactions were imposed using Abaqus general contact algorithm, employing a ``Hard Contact" model for the normal direction and a frictionless condition for the tangential direction to accurately capture scale substrate interactions. Case I serves as the baseline configuration for subsequent sROM analyses.

Finally, we note that this study does not yet include experimental confirmation. That omission is important, but it does not weaken the central mechanistic result, because the reduced-order model is validated against contact-resolved high-resolution FE simulations and because the main experimental obstacle is known - damping in common polymeric substrates can mask the very transitions predicted here. Still, the present analysis identifies the low-loss parameter regimes in which physical verification is most likely to succeed.

\begin{table}[h!]
\centering
\caption{FE model parameters and geometric dimensions for the biomimetic scale substrates in \textcolor{black}{FIG.~\ref{Fig3}} and \textcolor{black}{FIG.~\ref{Fig4}}}
\label{Tab:1}
\begin{tabular}{l c c c}
\hline
\textbf{Parameter} & \textbf{Case I} & \textbf{Case II} & \textbf{Unit} \\
\hline
\multicolumn{4}{c}{\textbf{Substrate Material Properties}} \\
\hline
Young’s modulus (\( E_B \)) & 10$^5$  &  10$^5$  & Pa \\
Poisson’s ratio (\( \nu_B \)) & 0.3 & 0.3 & - \\
mass density (\( \rho\)) & 1000  & 1000  & kg/m$^3$ \\
\hline
\multicolumn{4}{c}{\textbf{Substrate Geometry}} \\
\hline
length (\( L_B \)) & 1.00 & 0.25 & m \\
height (\( h_B \)) & 0.02 & 0.01 & m \\
width (\( b_B \))  & 0.02 & 0.02  & m \\
\hline
\multicolumn{4}{c}{\textbf{Scale Geometry}} \\
\hline
length (\( l \))  & 0.60 & 0.06  & m \\
pitch (\( d \))  & 0.06 & 0.01  & m \\
thickness (\( D \)) & 10$^{-4}$ & 2.6$\times$10$^{-4}$ & m \\
embedded length (\( L \)) & 0.005  & 0.005  & m \\
overlap ratio (\( \eta \)) & 10 & 6 & - \\
inclination angle (\( \theta_0 \)) & 2 & $\sim$ 0 & $^\circ$ \\
number of scales  & 17 & 25 & - \\
\hline
\multicolumn{4}{c}{\textbf{Loading Conditions}} \\
\hline
load amplitude (\( f_0 \))  & 5.7$\times$10$^{-3}$  & 5$\times$10$^{-2}$  & N/m \\
angular frequency (\( \omega \)) & 0.188 & 1.425 & rad/s \\
\hline
\end{tabular}
\end{table}

%%%%%%%%%%%%%%%%%%%%%%%%%%%%

%%%%%%%%%%%%%%%%%%%%%%%%%%%%%
\begin{figure*}
\includegraphics[width=1.0\linewidth]{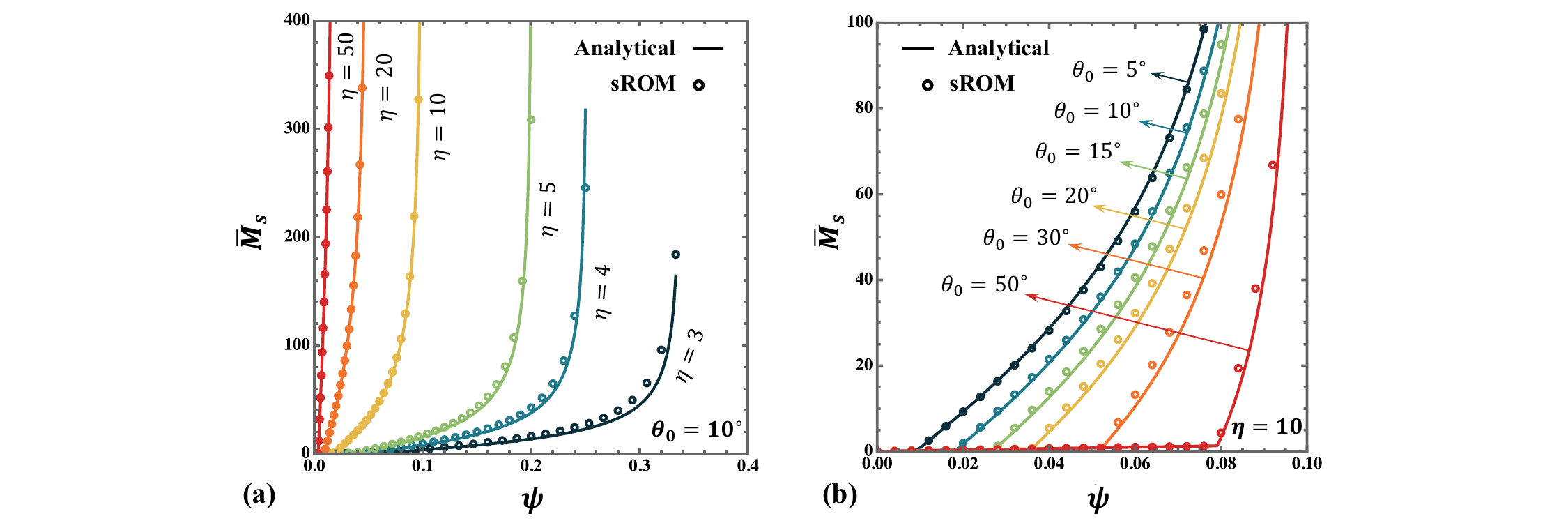}
\caption{\label{Fig2}\textbf{Normalized moment–curvature identification and surrogate accuracy.}
(a,b) Normalized moment $\bar{M}_{s}$ vs $\psi$ from the surrogate analytical model (solid lines) \textcolor{black}{\cite{ghosh2014contact}} versus the geometry-linked sROM (rings).
(a) $\theta_0=10^{\circ}$ with varying overlap ratio $\eta$. (b) $\eta=10$ with varying $\theta_0$. NRMSE$<5\%$ for $\lvert \kappa \rvert \le 0.9\,\kappa_{\mathrm{L}}$ in both $\eta$ and $\theta_0$ variations. Results correspond to Case I (see \textcolor{black}{TABLE \ref{Tab:1}}).}
\end{figure*}
%%%%%%%%%%%%%%%%%%%%%%%%%%%%%

%%%%%%%%%%%%%%%%%%%%%%%%%%%%%
\begin{figure*}

\includegraphics[width=1.0\linewidth]{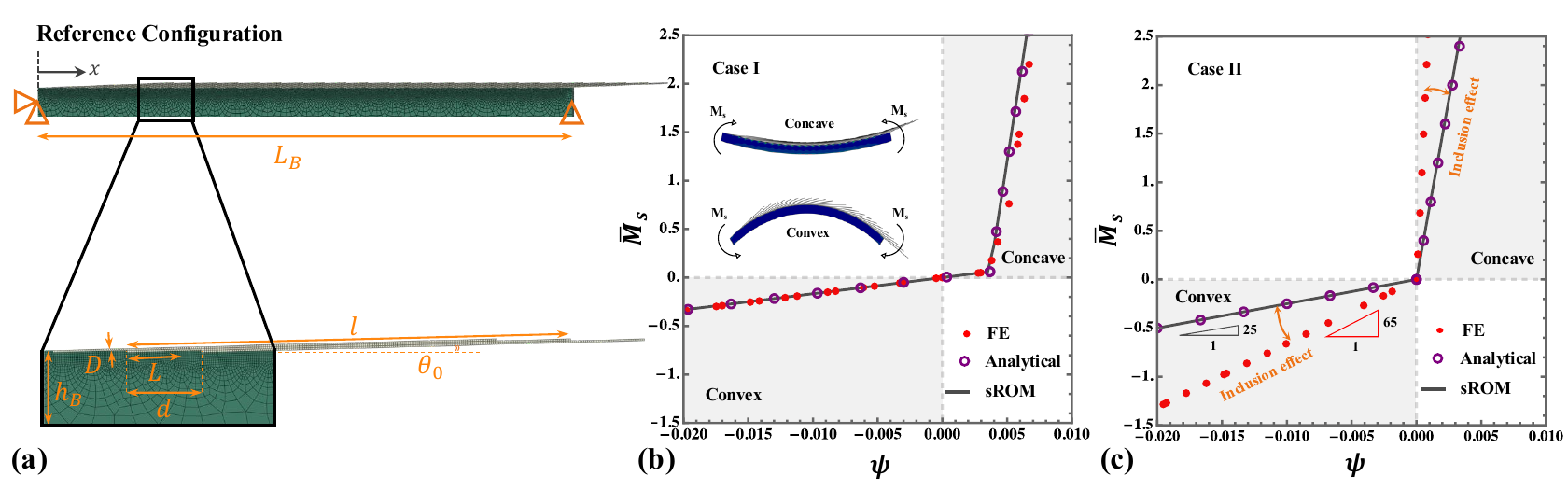}
\caption{\label{Fig3}  
\textbf{Normalized moment-curvature validation of sROM against FE and surrogate accuracy.} (a) FE model setup showing simply supported boundary conditions and geometric parameters. Quasi-static $\bar{M}_{s}-\psi$ responses for (b) Case I ($\eta=10, \theta_0=2^\circ$) and (c) Case II ($\eta=6, \theta_0\simeq 0^\circ$) (\textcolor{black}{see TABLE \ref{Tab:1}}): FE (red markers), analytical barrier (purple rings), and sROM (solid line). The response separates into convex ($\psi<0$) and concave ($\psi>0$) regions. sROM closely reproduces the FE response in Case I, whereas in Case II the FE response exhibits an increased slope on both convex and concave sides due to the inclusion effect arising from densely packed scales on the substrate \textcolor{black}{\cite{ebrahimi2019tailorable}}.}

\end{figure*}
%%%%%%%%%%%%%%%%%%%%%%%%%%%%%%

%%%%%%%%%%%%%%%%%%%%%%%%%%%%%
\begin{figure*}

\includegraphics[width=1.0\linewidth]{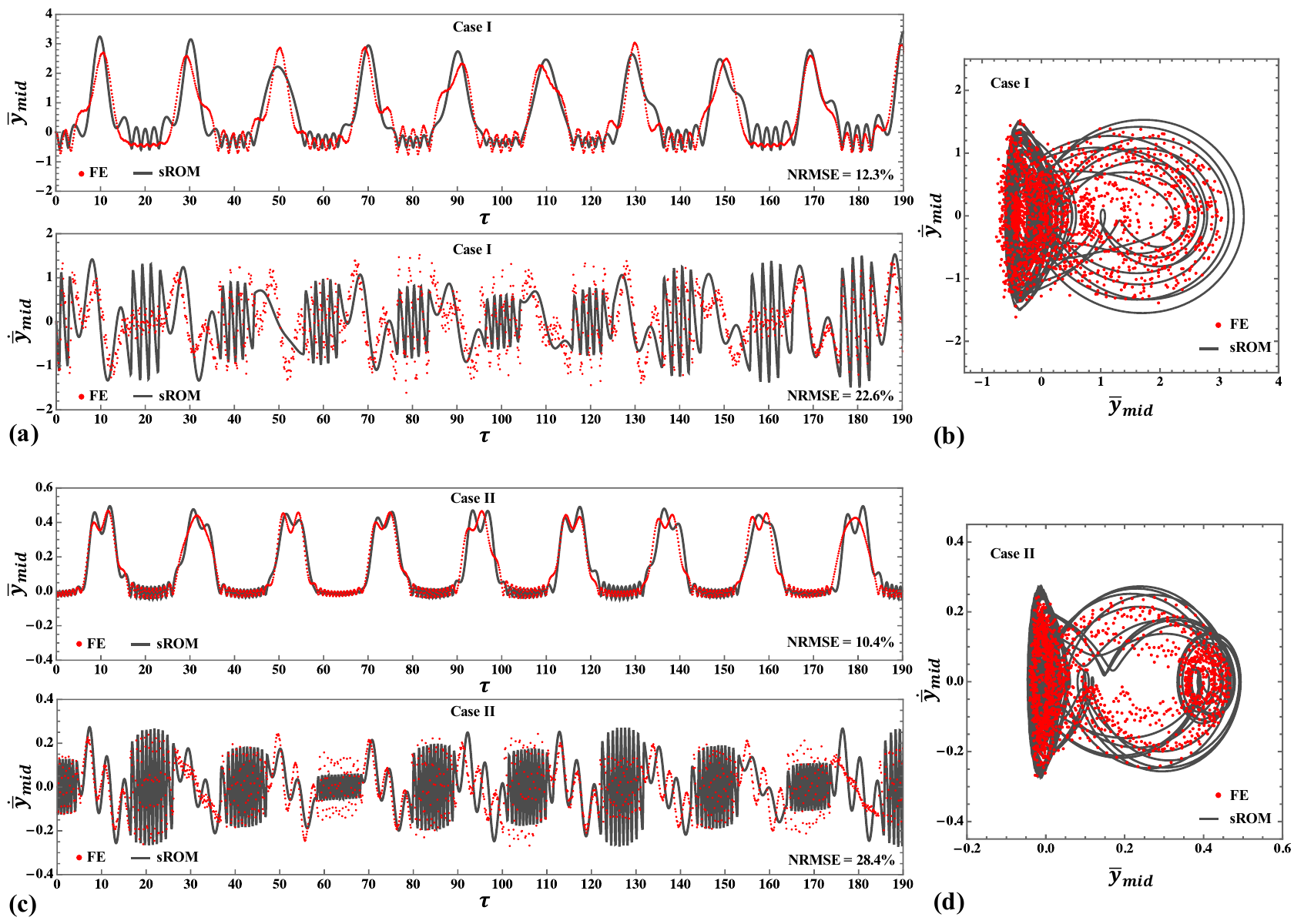}
\caption{\label{Fig4}  
\textbf{Dynamic validation of sROM versus FE.}
Forced response under harmonic excitation: (a) displacement (top), velocity (bottom), and (b) phase portrait for Case I ($\eta=10, \theta_0=2^\circ$); (c) displacement (top), velocity (bottom), and (d) phase portrait for Case II ($\eta=6, \theta_0\simeq 0^\circ$)(\textcolor{black}{see TABLE \ref{Tab:1}}). The sROM closely reproduces the FE time histories and phase-space trajectories in both cases. A damping ratio of $\delta=0.5\%$ is assumed in the sROM for both cases.}

\end{figure*}
%%%%%%%%%%%%%%%%%%%%%%%%%%%%%%

\begin{figure*}
\includegraphics[width=1\linewidth]{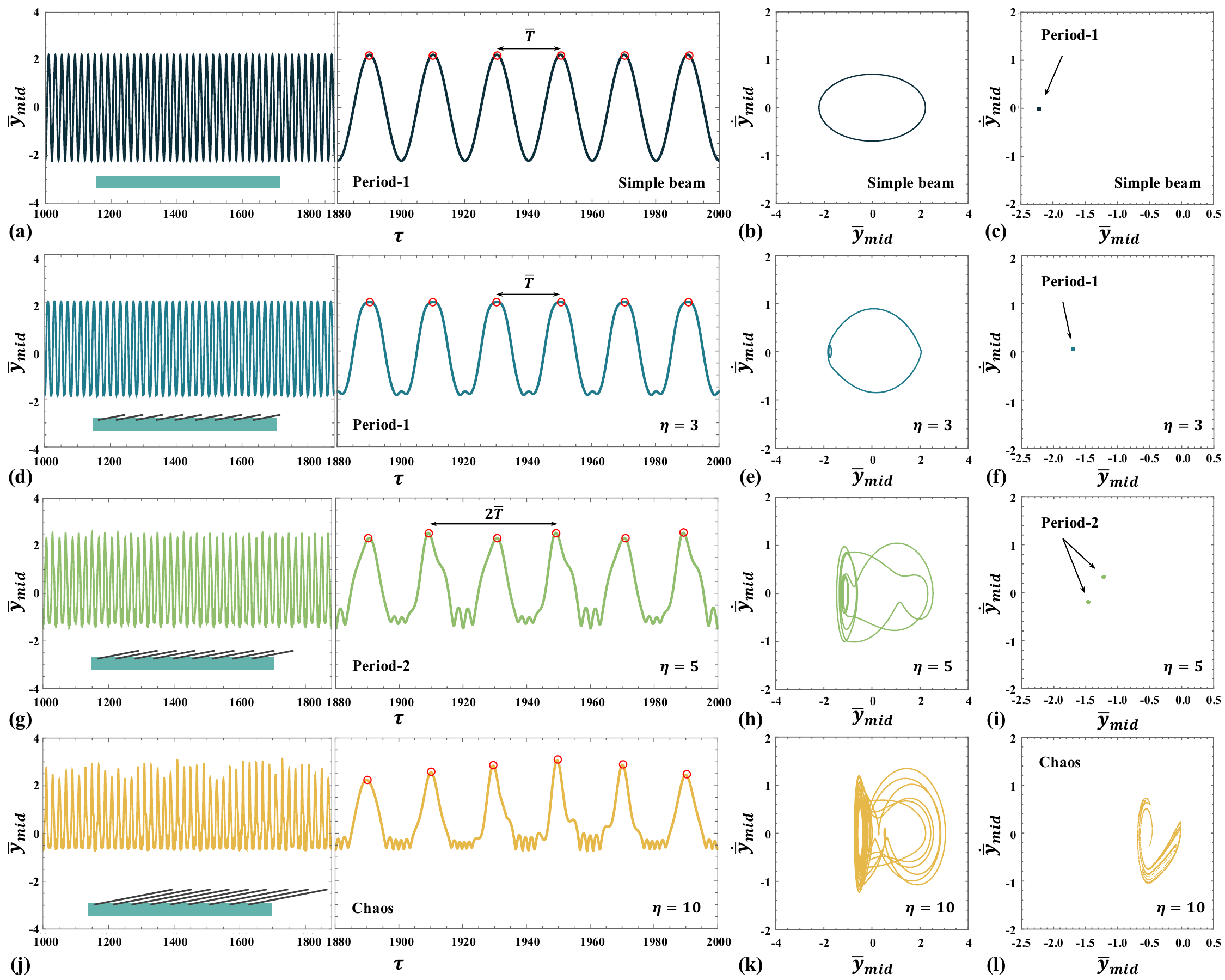}
\caption{\label{Fig5} 
\textbf{Tuning $\eta$ induces a bifurcation cascade from simple periodic motion to subharmonic and chaotic attractors.} (a) Simple beam: period-1 mid-span displacement. (b) Phase portrait: single closed orbit. (c) Poincaré section: single point. (d) $\eta=3$: period-1 response. (e) Phase portrait: bi-lobed limit cycle. (f) Poincaré section: single point. (g) $\eta=5$: period-2 response. (h) Phase portrait: doubled-period subharmonic orbit. (i) Poincaré section: two points (2-cycle). (j) $\eta=10$: chaotic response. (k) Phase portrait: chaotic attractor. (l) Poincaré section: smeared, fractal-like distribution with a broadband spectrum. All cases use $\theta_0=3^\circ$, $\delta=2\%$, $\gamma=2$, $\Omega=\pi/10$, and $\bar T=20$.}

\end{figure*}
%%%%%%%%%%%%%%%%%%%%%%%%%%%%%%

\section{Results and Discussion}
\label{Sec III}
 
We first validate the accuracy of our sROM by comparing the moment-curvature response across the critical geometrical system parameters - overlap $\eta$ and initial inclination $\theta_0$. This is shown in \textcolor{black}{FIG.~\ref{Fig2}(a,b)} where the sROM is shown to be exceedingly accurate across both of these parameters, capturing the characteristic three-regime response: linear, nonlinear, and finally the jammed (locked) state. The match is found to become more accurate at higher overlaps and lower initial inclinations. The obtained normalized root-mean-squared error (NRMSE) for $\lvert \kappa \rvert \le 0.9\,\kappa_{\mathrm{L}}$ remains below $5\%$, indicating an excellent fit and adequate predictive accuracy. The coefficients \( A \), \( \alpha \), and the exponent \( m \) defined in \textcolor{black}{Eq.~\ref{eq:4}}, are estimated as $A=K_B/2 $ per unit depth of the beam, $\alpha=0.95$, and $m=1.05$ by fitting the proposed function to previously high fidelity surrogate analytical models \textcolor{black}{\cite{ghosh2014contact}} for various combinations of \( \eta \) and \( \theta_0 \).

%%%%%%%%%%%%%%%%%%%%%%%%%%%%%
Next, we benchmark the fully resolved FE model and sROM against the analytical moment–curvature solution for quasi-static pure bending, as shown in \textcolor{black}{FIG.~\ref{Fig3}(a-c)}. The FE moment-curvature curve reproduces the expected tripartite structure predicted by analytical models \textcolor{black}{\cite{ghosh2014contact,ghosh2016frictional}}, a linear response prior to engagement, rapid strain stiffening once the scales impose unilateral contact, and a steep rise in moment as the curvature approaches the geometry-set locking barrier (jammed state). The normalized quasi-static moment-curvature responses for Case I and Case II are shown in \textcolor{black}{FIG. \ref{Fig3}(b,c)}, respectively, and both exhibit close agreement with analytical predictions. Small differences in slope from FE are due to the additional stiffness contributed by the embedded scale roots on the substrate side, which elevates the apparent tangent stiffness even at modest curvature \textcolor{black}{\cite{ebrahimi2019tailorable,dharmavaram2022coupled,sarkar2025bending}}. This effect is appreciable only in Case II, where $L/h_B > 0.25$ (see \textcolor{black}{FIG. \ref{Fig3}(c)}), and is amplified by the high packing density of the scales, which promotes early mechanical interaction and load transfer between adjacent scales. Notably, this inclusion effect manifests on both the convex and concave sides and can be incorporated into the sROM and analytical framework through an effective modulus, $E_B^{\mathrm{eff}}=\beta E_B$ \textcolor{black}{\cite{ebrahimi2019tailorable,dharmavaram2022coupled,sarkar2025bending}}, where $\beta$ represents the slope ratio and is approximately $\beta\simeq 65/25=2.6$ for Case II (see \textcolor{black}{FIG. \ref{Fig3}(c)}). In contrast, for Case I, no slope difference is observed, and the FE and analytical results are in excellent agreement \textcolor{black}{FIG. \ref{Fig3}(b)}.

We next excite the first-mode vibration using harmonic loading \( Q = f_0 \sin(\pi x / L_B) \cos(\omega t) \), and overlay the FE displacement and velocity responses with the sROM predictions over long time horizons \textcolor{black}{FIG.~\ref{Fig4}(a,c)}. The dynamic responses are organized separately for each configuration: \textcolor{black}{FIG. \ref{Fig4}(a)(top)} presents displacement, \textcolor{black}{FIG. \ref{Fig4}(a)(bottom)} shows velocity, and \textcolor{black}{FIG. \ref{Fig4}(b)} illustrates phase portraits for Case I, while \textcolor{black}{FIG. \ref{Fig4}(c,d)} shows the corresponding results for Case II. sROM shows good agreement with FE, with NRMSE of 12.3--10.4\% for displacement and 22.6--28.4\% for velocity across Cases~I--II. Notably, this level of agreement remains consistent across cases, indicating that the sROM accurately captures the contact-driven dynamics independent of geometric scaling. Further probing of the phase portraits \textcolor{black}{FIG.~\ref{Fig4}(b,d)} reveals that both the FE and sROM yield closely matched long-time orbits under identical excitation, as evidenced by both phase portraits and stroboscopic samples. This agreement persists across both configurations, as seen for Case I and Case II. Two additional features are worth noting: (i) a clear bias in the phase portrait, reflecting the asymmetric restoring law introduced by one-sided engagement; and (ii) a slow amplitude drift near the operating boundary, visible in the velocity trace near the edges of the operating window. This slight, slow ``breathing" of the velocity amplitude appears near the regime boundary, consistent with a near-contact event. Together, these observations justify using the sROM to chart the dynamics over parameter space at a fraction of the FE cost while retaining contact-driven mechanics. The success of our sROM enables high-credibility exploration of bifurcations governed purely by unilateral contact over a broad parameter space, without invoking large deflections or material dissipation.\\

%%%%%%%%%%%%%%%%%%%%%%%%%%%%%
\begin{figure*}
\includegraphics[width=1\linewidth]{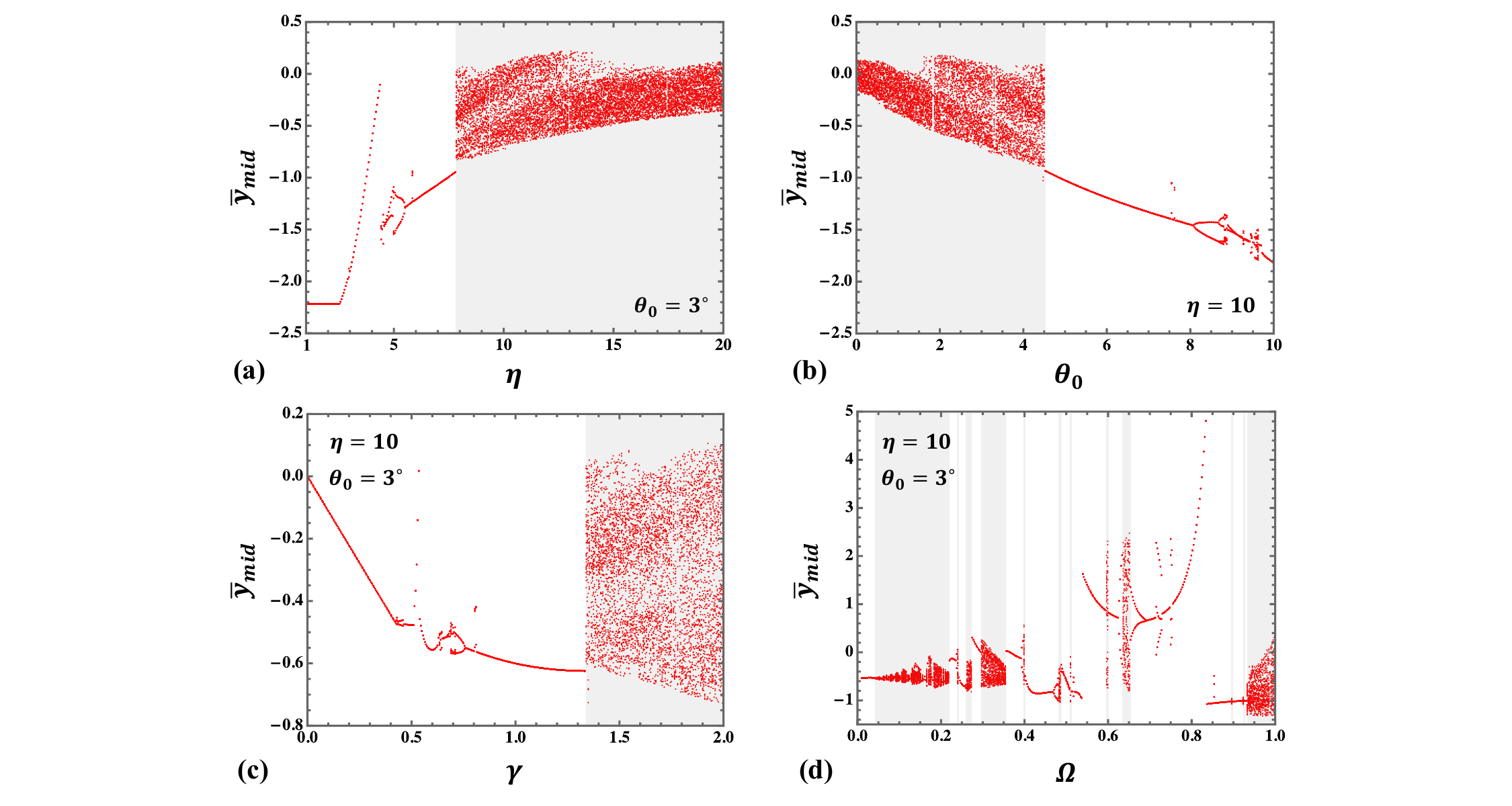}
\caption{\label{Fig6}
\textbf{Scale geometry ($\eta$ and $\theta_0$) tunes the onset and extent of chaos, while forcing amplitude $\gamma$ and frequency $\Omega$ select dynamical regimes.}
Bifurcation diagrams: displacement versus (a) $\eta$ for $\theta_0=3^\circ$, (b) $\theta_0$ for $\eta=10$, (c) $\gamma$ and (d) $\Omega$. Dark, dense bands mark a chaotic response. Fixed $\delta = 2\%$ for all, $\Omega = \pi/10$ for (a,b,c), and $\gamma=2$ for (a,b,d).}
\end{figure*}
%%%%%%%%%%%%%%%%%%%%%%%%%%%%%%

%%%%%%%%%%%%%%%%%%%%%%%%%%%%%

\begin{figure*}
\includegraphics[width=0.78\linewidth]{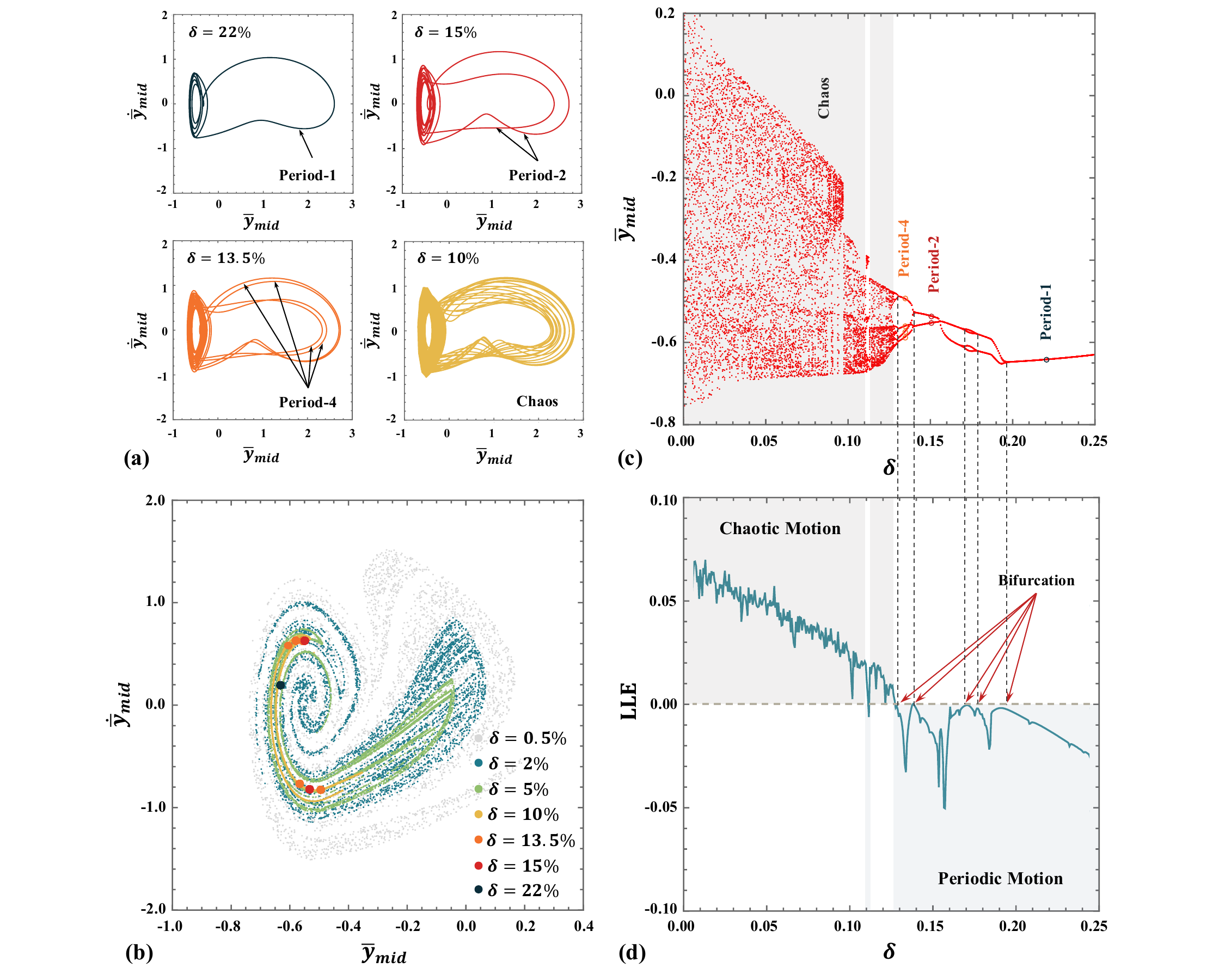}
\caption{\label{Fig7} 
\textbf{Damping suppresses the cascade.} (a) Phase portraits for $\delta \in \{10,\,13.5,\,15,\,22\}\%$: period-doubling route quenched by damping. (b) Poincaré sections for $\delta \in \{0.5,\,2,\,5,\,10,\,13.5,\,15,\,22\}\%$: emergence and splitting of periodic attractors en route to chaos. (c) Bifurcation diagram (displacement versus $\delta$): transition from period-$1 \rightarrow 2 \rightarrow 4$ $\rightarrow$ chaotic windows. (d) the largest Lyapunov exponent (LLE) versus $\delta$: LLE $>$ 0 in chaotic windows, returning to LLE $\le$ 0 as damping increases. Fixed $\eta = 10$, $\theta_0 = 3^\circ$, $\gamma = 2$, $\Omega = \pi/10$.}
\end{figure*}
%%%%%%%%%%%%%%%%%%%%%%%%%%%%%%

%%%%%%%%%%%%%%%%%%%%%%%%%%%%

We now delve into the effects of increasing $\eta$ on the nonlinear dynamics of the problem using time-history plots, phase portraits, and Poincaré maps. Poincaré maps are constructed by stroboscopically sampling the response at the (dimensionless) forcing period $\bar{T}=2\pi/\Omega$, reducing the continuous dynamics to a discrete representation that reveals the geometric structure of periodic, multi-periodic, and chaotic attractors and facilitates identification of chaotic dynamics. With $\eta=0$ (no scale), the response is harmonic: a single closed orbit in phase space and a single Poincaré point, \textcolor{black}{FIG.~\ref{Fig5}(a–c)}. Introducing unilateral contact ($\eta=3$) preserves period-1 but distorts the waveform and shifts the orbit off the origin, consistent with an asymmetric restoring law, \textcolor{black}{FIG.~\ref{Fig5}(d–f)}. At $\eta=5$, the system exhibits a period-2 orbit (two Poincaré points and a double temporal period, \textcolor{black}{FIG.~\ref{Fig5}(g–i)}). Further increase to $\eta=10$ yields a diffuse Poincaré set and a complex phase portrait, indicating a chaotic attractor, \textcolor{black}{FIG.~\ref{Fig5}(j–l)}. 
Mechanistically, increasing the geometrically tunable parameter $\eta$, which encodes the architecture of the fish-scale system, sharpens the post-contact restoring law and triggers a discontinuity-induced route to complex dynamics. As $\eta$ increases, higher-period response windows emerge and subsequently give way to chaos, all at modest excitation amplitudes and without invoking large deflections. These results demonstrate that contact engagement governed purely by geometric design provides a direct architectural pathway for selecting between narrow-band periodic responses and broadband chaotic regimes in an otherwise lightly excited beam, thereby enabling geometry-controlled tuning of the system response spectrum.

%%%%%%%%%%%%%%%%%%%%%%%%%%%%
Next, we examine how increasing the geometric parameters of the scales ($\eta,\theta_0$) and forcing parameters ($\gamma,\Omega$) reorganize the system response using bifurcation plots. Bifurcation plots are constructed by systematically varying a control parameter and recording the Poincaré-sampled response at each parameter value, thereby revealing qualitative changes in the stroboscopic dynamics. This approach enables identification of transitions between periodic, multi-periodic, and chaotic regimes, as well as routes to chaos. For $\theta_0=3^\circ$, the dynamics follow a familiar progression: a stable period-1 response at low $\eta$ that transitions through higher-period windows and intermittent complex behavior as $\eta$ increases \textcolor{black}{FIG.~\ref{Fig6}(a)}. 
For $\eta=10$, as $\theta_0$ decreases, engagement occurs early and the extent of the nonlinear region increases. This greater injection of nonlinearity can lead to stronger contact-induced instability, whereas larger $\theta_0$ favors more regular, low-period motion \textcolor{black}{FIG.~\ref{Fig6}(b)}.

Importantly, these behaviors shown in \textcolor{black}{FIG.~\ref{Fig6}(a,b)} reflect the fact that both $\theta_0$ and $\eta$ are geometric parameters that jointly regulate the post-engagement stiffness as well as the extent of the nonlinear range of the biomimetic system. Changing either parameter controls not only the initiation but also the extent of complex dynamics, with $\eta$ in particular sharpening the restoring law. These results demonstrate that post-engagement nonlinearity provides a purely architectural mechanism for programming the phase-space structure of fish-scale beams. Complex and chaotic oscillations can thus be induced, advanced, or suppressed through geometric design alone without reliance on large excitation amplitudes or material dissipation.

We now probe the effect of load amplitude $\gamma$ and frequency ratio $\Omega$ on the dynamical behavior of the fish-scale substrate by considering $\eta=10$ and $\theta_0=3^\circ$. In amplitude sweeps of $\gamma$, the bifurcation plot in \textcolor{black}{FIG.~\ref{Fig6}(c)} shows that, as $\gamma$ increases, scale engagement is eventually triggered, leading to a first bifurcation and a slight deviation from periodic response. With further increases in $\gamma$, the system begins to settle into alternating states of complex and simple oscillations, eventually leading to a dense bifurcation cloud indicating a transition to chaos. On the other hand, frequency sweeps of $\Omega$, \textcolor{black}{FIG.~\ref{Fig6}(d)}, reveal alternating bands of periodic and chaotic response as $\Omega$ detunes across resonance, forming tongue-like structures with narrow periodic islands embedded within broader chaotic seas. In particular, the fine vertical striations indicate intermittent windows. 
Together, these trends indicate that increasing forcing amplitude $\gamma$ primarily broadens the response, increasingly bringing the system closer to bifurcation (\textcolor{black}{FIG.~\ref{Fig6}(c)}), whereas the excitation frequency $\Omega$ reorganizes response windows through resonance-tongue modulation (\textcolor{black}{FIG.~\ref{Fig6}(d)}).\\

%%%%%%%%%%%%%%%%%%%%%%%%%%%%%%

\begin{figure*}
\includegraphics[width=1.0\linewidth]{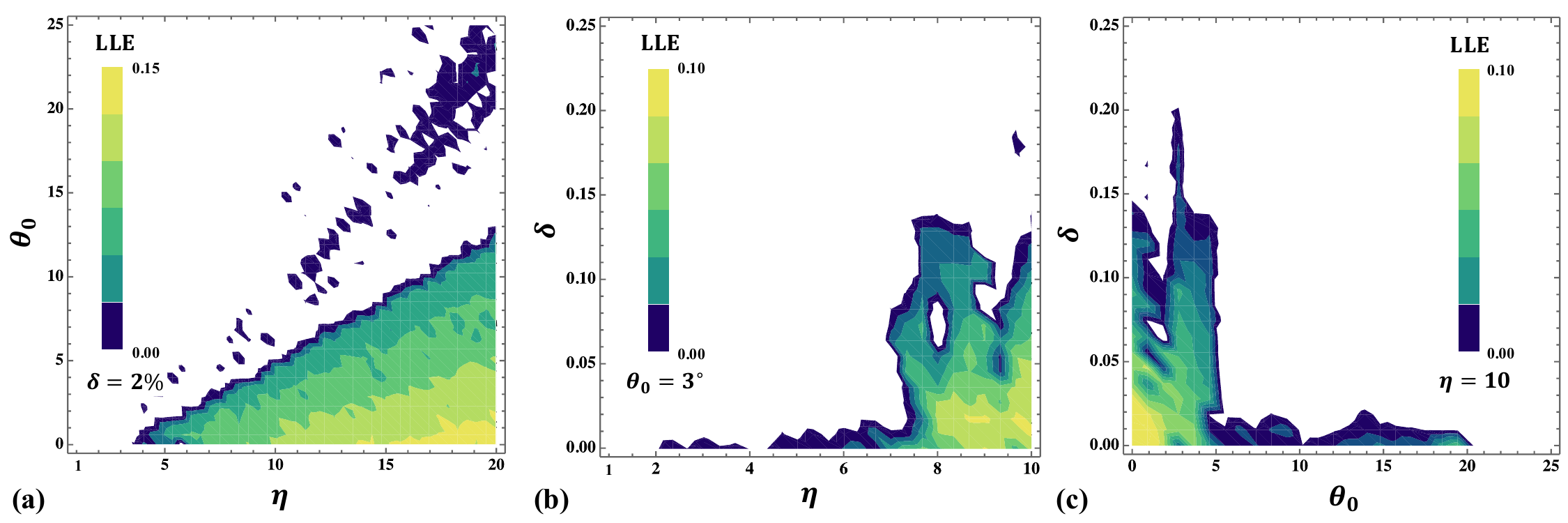}
\caption{\label{Fig8} 
\textbf{Chaos maps reveal the influence of geometry and damping in the biomimetic scale substrate.} The largest Lyapunov exponent (LLE) distributions are shown in (a) $\eta$–$\theta_0$, (b) $\eta$–$\delta$, and (c) $\theta_0$–$\delta$ parameter spaces. Color-coded values of the LLE identify transitions between periodic, multi-periodic, and chaotic regimes. Fixed $\gamma=2$ and $\Omega=\pi/10$.}
\end{figure*}

%%%%%%%%%%%%%%%%%%%%%%%%%%%%%%

Damping is another critical variable that reshapes the asymmetric dynamics by opening or closing windows created by engagement (high damping would suppress all chaos). In fish-scale systems, damping can be highly complex and non-intuitive, leading to emergent phenomena such as viscous damping even when only Coulomb friction is present \textcolor{black}{\cite{ali2019frictional}}. Overall damping, like stiffness, can be significantly tuned by geometry alone. For modest excitation frequencies, damping of various types can be approximated using an effective damping ratio $\delta$ \textcolor{black}{\cite{ebrahimi2023material}}. As the damping ratio $\delta$ increases, the stroboscopic response contracts from multi-period sets to a single period-1 orbit, whereas reducing $\delta$ reverses this path, \textcolor{black}{FIG.~\ref{Fig7}(a)}. In this figure, the phase portraits clearly show the sequence: period-1 at $\delta=22\%$, period-2 at 15$\%$, period-4 at 13.5$\%$, and a chaotic attractor by 10$\%$. The Poincaré overlay in \textcolor{black}{FIG.~\ref{Fig7}(b)} stacks these regimes at fixed strobe phase and reveals a smooth contraction of the invariant set as damping grows. The bifurcation diagram (\textcolor{black}{FIG.~\ref{Fig7}(c)}) demonstrates a classic period-doubling progression with intermittent narrow periodic windows. Further evidence is established by plotting the dependence of the largest Lyapunov exponent (LLE) on damping. The LLE provides a quantitative measure of the average rate at which nearby trajectories in the system phase space diverge (or converge) over time. For systems governed by nonlinear ODEs, a positive LLE indicates chaotic behavior, whereas zero or negative values correspond to multi-periodic or periodic dynamics (see \textcolor{black}{\ref{SM2}} for more details). For this system, the LLE crosses zero at the same point as the critical points in bifurcation plots (\textcolor{black}{FIG.~\ref{Fig7}(d)}) and remains positive throughout the chaotic band. It decreases monotonically as $\delta$ is increased. Together, these diagnostics confirm that $\delta$ serves as a practical geometric and interfacial control parameter. Damping thus tunes access to complexity. It can postpone or suppress contact-induced cascades, providing a straightforward knob to select narrow-band periodic responses or broadband chaotic regimes within the same architected beam. Given the multiple contributing factors to damping, geometry, external fluids, and interfacial properties, it offers one of the most versatile levers for chaos modulation.\\

\textcolor{black}{FIG.~\ref{Fig8}} maps LLE across complementary geometric–dissipative parameter spaces, quantifying how contact architecture governs the transition between periodic and chaotic dynamics. 
In \textcolor{black}{FIG.~\ref{Fig8}(a)}, the $(\eta,\theta_0)$ map at fixed damping $\delta=2\%$ reveals a wedge-shaped region of positive LLE, indicating that chaos emerges within a structured geometric corridor rather than uniformly across parameter space. 
Physically, increasing the scale overlap ratio $\eta$ intensifies contact interactions and sharpens the post-engagement restoring law, while decreasing the initial inclination $\theta_0$ advances the onset of scale engagement (earlier contact). The coupled increase in $\eta$ and decrease in $\theta_0$ promotes earlier and stronger engagement. This greatly increases the overall nonlinear energy transfer during oscillations, amplifying the chaotic tendencies of the system.

%%%%%%%%%%%%%%%%%%%%%%%%%%%%%%%

\begin{figure*}
\includegraphics[width=1.0\linewidth]{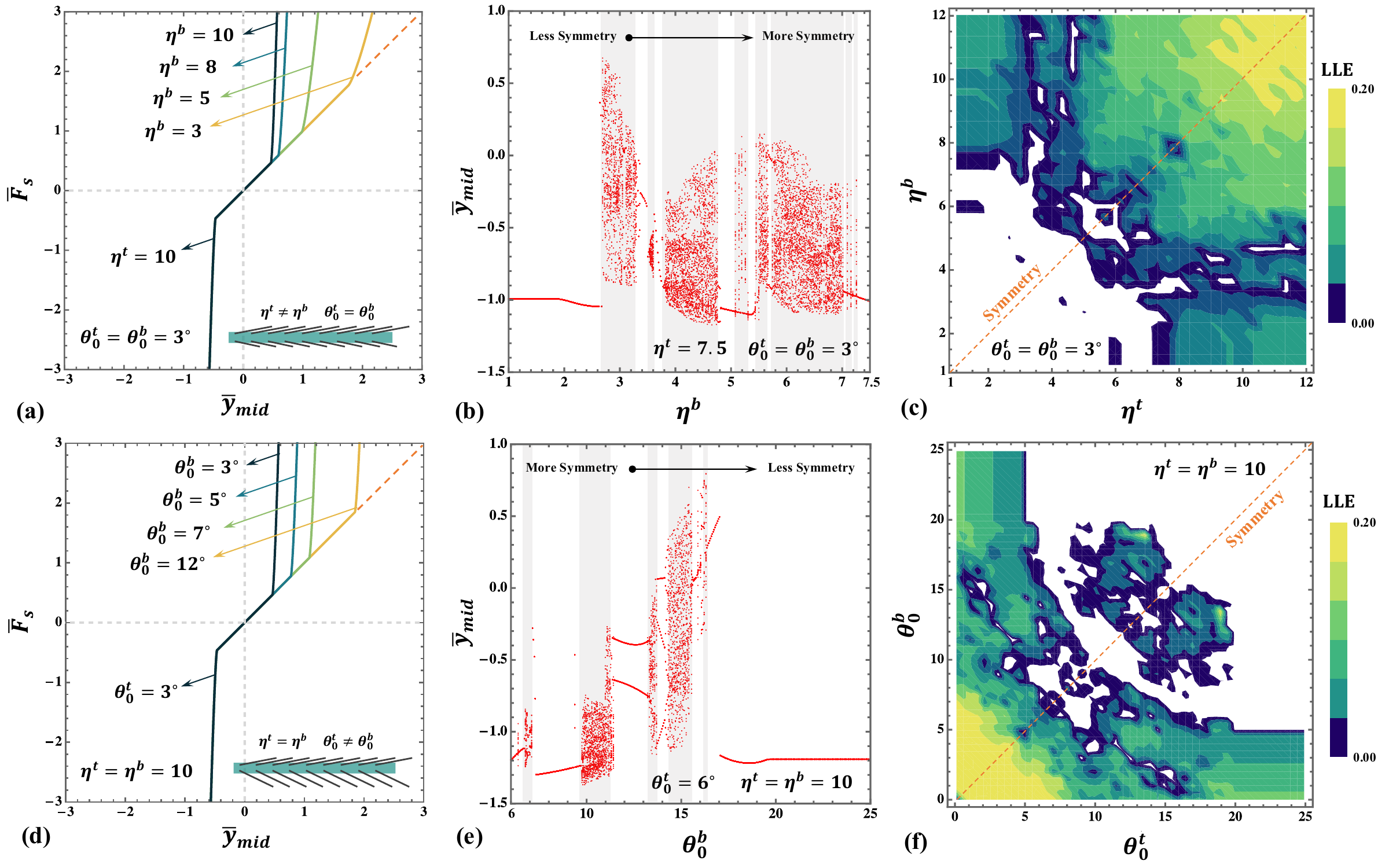}
\caption{\label{Fig9} 
{\textbf{Asymmetry from one-sided textures biases phase space and regulates chaotic dynamics.}
(a) Normalized force–displacement curves for varying $\eta^{b}$, showing the transition from strongly asymmetric to symmetric configurations. (b) Bifurcation diagram of normalized mid-span displacement versus $\eta^{b}$; chaotic regions are highlighted. (c) Chaos map in the $\eta^t-\eta^b$ parameter space, where color-coded values of the largest Lyapunov exponent (LLE) identify transitions between periodic, multi-periodic, and chaotic regimes. (d) Normalized force–displacement curves for varying $\theta_0^{b}$, showing the transition from strongly asymmetric to symmetric configurations. (e) Bifurcation diagram of normalized mid-span displacement versus $\theta_0^{b}$, showing the role of symmetry breaking in selecting dynamical regimes; chaotic regions are highlighted. (f) Chaos map in the $\theta_0^t-\theta_0^b$ parameter space, where color-coded values of the LLE identify transitions between periodic, multi-periodic, and chaotic regimes. Superscripts $t$ and $b$ denote the top and bottom sides of the biomimetic scale substrate. Fixed $\delta = 2\%$, $\gamma = 2$, and $\Omega = \pi/10$.}}
\end{figure*}

%%%%%%%%%%%%%%%%%%%%%%%%%%%%%%%

In contrast, \textcolor{black}{FIG.~\ref{Fig8}(b,c)} isolates the role of dissipation by mapping the ($\eta,\delta$) and ($\theta_0,\delta$) respectively. Chaotic regions concentrate at low damping, large overlap, and also low scale inclination, while increasing $\delta$ progressively suppresses positive LLE, shrinking chaotic zones into narrower regions. This behavior reflects the competition between geometry-induced strain stiffening, which becomes active once contact is established, and damping, which attenuates the energy exchange associated with repeated contact interactions. Notably, moderate damping does not completely eliminate chaos when geometric stiffening is sufficiently strong, underscoring that dissipation completely forbids chaos only beyond a moderate range, leaving the possibility of chaotic vibrations for highly overlapped beams in a moderate damping regime. Taken together, these maps demonstrate that chaos in biomimetic scale substrates is primarily programmed by geometry, with $\eta$ and $\theta_0$ acting as architectural control parameters that structure the phase space, while damping serves as a secondary regulator below a threshold. Note that warm colors in the maps correspond to larger positive values of LLE, indicating stronger chaotic behavior and higher dynamical irregularity, whereas cooler colors denote weakly nonlinear or periodic responses with limited complexity.\\

%%%%%%%%%%%%%%%%%%%%%%%%%%%%%%%

Another distinctive feature of biomimetic scale substrates is that asymmetry is built directly into the architecture through unequal scale density or inclination on the two sides of the beam. We first consider scale overlap asymmetry, quantified by the overlap ratios on the top and bottom surfaces, $\eta^t$ and $\eta^b$. The resulting force-displacement curves (force from \textcolor{black}{Eq.~\ref{eq:9}} and displacement is of beam midpoint) in \textcolor{black}{FIG.~\ref{Fig9}(a)} show that asymmetry immediately breaks the antisymmetry of the restoring response as the curve acquires an offset about the origin and develops unequal slopes on the two half-cycles. As $\eta^b$ approaches $\eta^t$, this bias weakens and the restoring law recovers toward a more symmetric form. The dynamical consequence is shown in the bifurcation plot of \textcolor{black}{FIG.~\ref{Fig9}(b)}. Here, the key observation is that increasing symmetry does not reduce chaotic behavior. In fact, it is at high asymmetries that chaos seems to disappear. If one side were too sparse, the scales would not engage on that side. This would lead to zones of horizontal straight lines on this bifurcation behavior ($\eta$ -independence on one side), thereby creating an effectively one-sided system studied earlier. 
The interplay of chaos and symmetry is more complex and it requires spanning the chaos landscape using LLE maps, \textcolor{black}{FIG.~\ref{Fig9}(c)}. This picture shows that, in general, symmetric configurations can exhibit chaos even at relatively low scale overlaps when compared to asymmetric counterparts. To ensure chaotic vibration in asymmetric distribution, the overlap must be considerably higher. As overlap increases, the most intense chaotic instabilities start to become more concentrated around the line of symmetry. A closely related picture emerges for angular asymmetry, in which the two surfaces carry identical overlap but different inclination angles, $\theta_0^t$ and $\theta_0^b$. The force-displacement curves in \textcolor{black}{FIG.~\ref{Fig9}(d)} show that increasing $\theta_0^b$ progressively delays engagement on the lower side, thereby amplifying the restoring bias and reducing the contribution of the bottom texture to the cycle. The bifurcation plot in \textcolor{black}{FIG.~\ref{Fig9}(e)} again shows that asymmetry does not act as a simple destabilizer. The strongest chaotic response occurs near the more symmetric, low-angle configurations, whereas larger angular mismatch delays chaotic onset. The corresponding LLE map in \textcolor{black}{FIG.~\ref{Fig9}(f)} makes this structure explicit. Chaos is favored when both angles are small and comparable, reflecting strong bilateral engagement, while increasing $\theta_0^b$ delays or suppresses lower-side contact. The extended horizontal clusters at large $\theta_0^b$ indicate that the bottom scales are dynamically inactive, so the system again reduces to an effectively one-sided architecture. Taken together, \textcolor{black}{FIG.~\ref{Fig9}(a-f)} shows that in these systems broken symmetries can have a stabilizing effect rather than the usual destabilizing effect common in many dynamical systems. 
Overall geometric symmetry can act a regulator that biases the restoring law, separates engagement thresholds across the cycle, delays chaotic onset, and reorganizes the phase space into chaotic bands, periodic islands, and can effectively make one side inactive from lack of engagement. Both overlap ratio asymmetry and inclination asymmetry act as independent geometric routes to broken symmetry $\eta,\theta_0$. Yet their dynamical consequences are strongly analogous, producing qualitatively similar restructuring of the bifurcation and chaos maps.

\section{Conclusions}
\label{Sec IV}

We have shown that biomimetic scale substrates constitute a distinct class of contact-governed oscillators in which ordinary beam bending is converted into a complex oscillator whose dynamics are governed by unilateral engagement and progressive jamming. A singular reduced-order model (sROM), derived from continuum mechanics and validated against contact-resolved finite element (FE) simulations, is highly effective in capturing this mechanism while preserving the essential physics of pre-engagement bending, post-engagement hardening, and approach to locking. Within this framework, the route to chaos is organized and interpretable. The overlap and scale inclination regulate the onset and extent of post-engagement nonlinearity, damping contracts or suppresses chaotic windows, and top-bottom asymmetry breaks restoring antisymmetry, arresting chaotic onset. These findings matter for two reasons. First, they show that deterministic chaos in a structural substrate need not rely on large deflection or constitutive complexity, emerging at modest amplitudes from geometry alone. Second, they recast overlapping surface architecture from a passive covering to an active dynamical mechanism. In that sense, the fish-scale substrate studied here is not merely a biomimetic curiosity but also a canonical member of a broader family of textured beams and ribbons in which contact geometry can be used to select periodic, multi-periodic, or chaotic motion by design. The central implication of this work is therefore not only that chaos can arise in such systems, but that it can be advanced, delayed, reorganized, or suppressed through architecture. Contact, in these substrates, is not merely something to resolve numerically; it is a structural mechanism for programming oscillatory response.

\begin{acknowledgments}
This work was supported by the United States National Science Foundation’s Civil, Mechanical, and Manufacturing Innovation, Grant No. $2028338$.
\end{acknowledgments}

\section*{Supplementary material}

\setcounter{section}{0}
\setcounter{equation}{0}
\renewcommand{\theequation}{S\arabic{equation}}
\renewcommand{\thesection}{SM\arabic{section}}

\refstepcounter{section}
\subsection*{\thesection. Nonlinear Moment–Curvature Relation and Governing Equations}
\label{SM1}

The moment–curvature relation derived using the singular reduced-order model (sROM) is expressed as follows:

\begin{equation}
\label{SMeq:1}
M_{s}(y'')=B y''+\tilde{H}(y''-y''_e) A\eta^m  \tan\left( \frac{\pi}{2} \alpha \frac{y''-y''_e}{y''_L-y''_e} \right)
\end{equation}

The governing nonlinear partial differential equation (PDE) describing the dynamics of a simply-supported biomimetic scale substrate is given by:

\begin{equation}
\label{SMeq:2}
\rho S\ddot{y}+c_d\dot{y}
+\frac{\partial^2 M_{s}(y'')}{\partial x^2}
= f_0\sin(\pi x/L_B)\cos(\omega t)
\end{equation}

The PDE above can be cast in nondimensional form by identifying two intrinsic geometric length scales: the beam length $L_B$ characterizing horizontal length scales and the beam height $h_B$, representing vertical length scales. We define the ratio of these scales as $\zeta=h_B/L_B$. A natural time scale is extracted from the fundamental bending frequency of the underlying simply-supported elastic substrate, $\omega_0 = \pi^2 \sqrt{B/\rho S L_B^4}$ which enables the following nondimensional variables $\bar{x}=x/L_B$,  $\bar{d}=d/L_B$,  $\bar{y}=y/h_B$, $\tau=\omega_0 t$, $\Omega=\omega/\omega_0$. Using these nondimensional parameters, the moment–curvature relation converts to the following nondimensional form:

\begin{equation}
\label{SMeq:3}
\bar{M}_{s}(\bar{y}'')=\zeta \bar{y}''+\tilde{H}(\bar{y}''-\bar{y}''_e) \bar{K}\eta^m  \tan\left( \frac{\pi}{2} \alpha \frac{\bar{y}''-\bar{y}''_e}{\bar{y}''_L-\bar{y}''_e} \right)
\end{equation}

where nondimensional bending moment is defined as $\bar{M}_s=M_sL_B/B$, and the nondimensional stiffness is defined as $\bar{K}=A L_B/B$. Also $\bar{y}''=\psi/\zeta\bar{d}$. Accordingly, the governing PDE reduces to the following nondimensional representation:

\begin{equation}
\label{SMeq:4}
\ddot{\bar{y}}+\delta\dot{\bar{y}}
+\frac{1}{\pi^4\zeta}\frac{\partial^2 \bar{M}_{s}(\bar{y}'')}{\partial \bar{x}^2}
= \gamma\sin(\pi \bar{x})\cos(\Omega \tau)
\end{equation}

where $\delta = c_d / \rho S \omega_0$ and $\gamma = f_0 / \rho S h_B \omega_0^2$. To simplify the nondimensional nonlinear PDE above and eliminate the second spatial derivative of the bending moment, we employ a Galerkin weak-form approach. The equation is projected onto a weighting function chosen to be identical to the fundamental first eigenmode of the beam ($\phi_1(\bar{x})=\sin(\pi \bar{x})$). This choice inherently satisfies the boundary conditions (simply-supported) and eliminates the boundary terms arising from integration by parts. In addition, it yields a reduced-order representation in which the external forcing assumes a simplified modal form. The final Galerkin weak-form of the nonlinear PDE is therefore written as:

\begin{align}
\label{SMeq:5}
\int_{0}^{1} &\phi_1(\bar{x}) \ddot{\bar{y}} \, d\bar{x}
+ \delta \int_{0}^{1} \phi_1(\bar{x}) \dot{\bar{y}} \, d\bar{x}\notag\\
&+ \frac{1}{\pi^{4}\zeta} \int_{0}^{1} \phi_1{''}(\bar{x}) \bar{M}_s \, d\bar{x}
=
\gamma \cos(\Omega \tau)
\int_{0}^{1} \phi_1^{2}(\bar{x}) \, d\bar{x}
\end{align}

By employing separation of variables $\bar{y}=\phi_1(\bar{x}) Y(\tau)$, the nondimensional weak-form PDE is reduced to a nonlinear ordinary differential equation (ODE) governing the modal amplitude of the biomimetic scale substrate. This procedure yields the normalized nonlinear ODE in the following form:

\begin{equation}
\label{SMeq:6}
\ddot{Y}
+ \delta \dot{Y}
+ \frac{I_1}{\pi^{4} I_0} Y
+ \frac{\bar{K}\eta^{m}}{\pi^{4}\zeta I_0}
I_{NL}(Y(\tau),\eta,\theta_0,\zeta,\bar{d})
=
\gamma \cos(\Omega \tau)
\end{equation}

where $I_0$, $I_1$, and $I_{NL}$ are the mode-shape integrals and are defined as following:

\begin{align}
\label{SMeq:7}
&I_0 = \int_{0}^{1} \phi_1^2(\bar{x}) \, d\bar{x}
= \frac{1}{2}, \quad I_1 = \int_{0}^{1} \phi_1{''}^2(\bar{x}) \, d\bar{x}
= \frac{\pi^4}{2}, \notag\\[8pt] 
&I_{NL} = \int_{0}^{1}
\phi_1{''}(\bar{x})\,
\tilde{H}\!\left(\phi_1{''}(\bar{x}) Y(\tau) - \bar{y}_e'' \right)
\times \notag\\ 
&\qquad\qquad\qquad\qquad\qquad\quad \tan\!\left(
\frac{\pi}{2}\alpha
\frac{\phi_1{''}(\bar{x}) Y(\tau) - \bar{y}_e''}
{\bar{y}_L'' - \bar{y}_e''}
\right)
\, d\bar{x}
\end{align}

This substitution guarantees a continuously differentiable restoring force, thereby improving numerical stability and enabling well-posed variational projection within the Galerkin framework. Specifically, the hyperbolic tangent function serves as a smooth approximation of the Heaviside operator $\tilde{H}(.)$.\\

Therefore, the final nonlinear ODE given below represents the reduced-order dynamics of the biomimetic scale substrate. The resulting equation is formally equivalent to a nonlinear spring-mass–damper (nSMD) system, in which the restoring force incorporates contact-induced geometric nonlinearities.

\begin{align}
\label{SMeq:8}
&\ddot{Y}(\tau)
+ \delta\,\dot{Y}(\tau)+\bar{F_s}=\gamma\,\cos\bigl(\Omega\,\tau\bigr) \notag\\[8pt]
&\bar{F_s}=Y(\tau)+\bar{F}_{NL}\bigl(Y(\tau), \bar{K}, \eta, \theta_0, \zeta, \bar{d}\bigr)
\end{align}

where $\bar{F}_{NL}= \frac{\bar{K}\eta^{m}}{\pi^{4}\zeta I_0}
I_{NL}$.

\refstepcounter{section}
\edef\currentlabel{SM\arabic{section}}

\subsection*{\currentlabel. Largest Lyapunov exponent}
\label{SM2}

The largest Lyapunov exponent (LLE) is a quantitative measure that evaluates the average exponential rate at which nearby trajectories in the system’s phase space diverge (or converge) over time. For systems governed by nonlinear ODEs, a positive LLE indicates chaotic behavior, while zero or negative values correspond to multi-periodic or periodic dynamics.

To compute the LLE, we consider the normalized and nondimensionalized nonlinear spring–mass–damper (nSMD) system that represents the dynamic behavior of a biomimetic scale substrate. Letting \( u = Y \) and \( v = \dot{Y} \), we define the state vectors as:

\begin{multline}
\mathbf{X} = 
\begin{pmatrix} 
u \\[6pt] 
v 
\end{pmatrix}, 
\quad
\dot{\mathbf{X}}=\mathbf{f}(\mathbf{X},\tau) = \begin{pmatrix} 
\dot{u} \\[6pt] 
\dot{v}
\end{pmatrix}\\
= \begin{pmatrix}
v \\[6pt]
-\delta v -  u -\bar{F}_{NL}(u) + \gamma \cos(\Omega \tau)
\end{pmatrix}
\end{multline}

Linearizing the system about a reference trajectory \( \mathbf{X}(\tau) \), we obtain the variational equations $
\tilde{\delta}\dot{\mathbf{X}} = D_{\mathbf{X}}\mathbf{f}(\mathbf{X}(\tau), \tau)\,\tilde{\delta}\mathbf{X}$ where $\tilde{\delta}$ denotes the variation operator, and the Jacobian matrix is given by:

\begin{equation}
D_{\mathbf{X}}\mathbf{f} =
\begin{pmatrix}
0 & 1 \\[6pt]
-1 - \frac{\partial \bar{F}_{NL}(u)}{\partial u}  & -\delta
\end{pmatrix}
\end{equation}

The LLE is computed by numerically evolving the perturbation vector \( \tilde{\delta}\mathbf{X} \), periodically rescaling its magnitude to a small value \( \epsilon \), and recording its growth over time. The final expression for the LLE is:

\begin{equation}
\text{LLE} = \lim_{N \to \infty} \frac{1}{N\,\Delta t} \sum_{p=1}^{N} \ln\left(\frac{\|\tilde\delta\mathbf{X}(t_p)\|}{\epsilon}\right)
\end{equation}

where \( N \) denotes the total number of renormalization iterations, \( t_p = p\,\Delta t \) represents the discrete sampling times, \( \Delta t \) is the fixed time interval between successive renormalizations, \( \|\cdot\| \) indicates the Euclidean norm of the perturbation vector, and \( \epsilon \) is the prescribed magnitude of the initial perturbation used for normalization. A positive LLE confirms the presence of chaos.

\bibliographystyle{apsrev4-2}
\bibliography{Main}% Produces the bibliography via BibTeX.

\end{document}